\documentclass[12 pt,twoside,aps,prd,amsmath,amssymb,
tightenlines,showpacs,showkeys,eqsecnum]{revtex4}
\usepackage{epsfig}
\usepackage{psfig}
\usepackage{bm}
\usepackage{graphicx}

\usepackage{hyperref}
\renewcommand{\cal}{\mathcal}

\newcommand {\ve}{\varepsilon}

\newcommand {\prm}{\prime}

\newcommand {\cL}{\cal L}
\newcommand {\G}{\Gamma}

\def \myfigs #1#2#3#4#5#6#7#8
{\begin{figure}[ht]
    \begin{center}
        \includegraphics[width=#2 \textwidth]{#1.eps}
        \hfill
        \includegraphics[width=#6 \textwidth]{#5.eps}
        \parbox[t]{#4\textwidth}{\caption {#3}\label{#1}}
        \hfill
        \parbox[t]{#8\textwidth}{\caption {#7}\label{#5}}
    \end{center}
\end{figure} }

\def \myfigures #1#2#3#4#5#6#7#8
{\begin{figure}[ht]
    \begin{center}
        \includegraphics[width=#2 \textwidth, angle=-90]{#1.eps}
        \hfill
        \includegraphics[width=#6 \textwidth, angle=-90]{#5.eps}
        \parbox[t]{#4\textwidth}{\caption {#3}\label{#1}}
        \hfill
        \parbox[t]{#8\textwidth}{\caption {#7}\label{#5}}
    \end{center}
\end{figure} }

\def\myfigure #1#2#3#4
{\begin{figure}[ht]\begin{center}
\includegraphics[width=#2 \textwidth, angle = -90]{#1.eps}
\parbox[t]{#4\textwidth}{\caption{#3}\label{#1}}
\end{center}\end{figure}}

\begin{document}
\title{Bianchi type I universe with viscous fluid}
\author{Bijan Saha}
\affiliation{Laboratory of Information Technologies\\
Joint Institute for Nuclear Research, Dubna\\
141980 Dubna, Moscow region, Russia} \email{saha@thsun1.jinr.ru}
\homepage{http://thsun1.jinr.ru/~saha/}
\date{\today}

\begin{abstract}
We study the evolution of a homogeneous, anisotropic Universe
given by a Bianchi type-I cosmological model filled with viscous
fluid, in the presence of a cosmological constant $\Lambda$. The
role of viscous fluid and $\Lambda$ term in the evolution the BI
space-time is studied. Though the viscosity cannot remove the
cosmological singularity, it plays a crucial part in the formation
of a qualitatively new behavior of the solutions near singularity.
It is shown that the introduction of the $\Lambda$ term can be
handy in the elimination of the cosmological singularity. In
particular, in case of a bulk viscosity, it provides an
everlasting process of evolution ($\Lambda < 0$), whereas, for
some positive values of $\Lambda$ and the bulk viscosity being
inverse proportional to the expansion, the BI Universe admits a
singularity-free oscillatory mode of expansion. In case of a
constant bulk viscosity and share viscosity being proportional to
expansion, the model allows oscillatory mode accompanied by an
exponential growth even with a negative $\Lambda$. Space-time
singularity in this case occurs at $t \to -\infty$.
\end{abstract}



\maketitle

\bigskip


\section{Introduction}

The investigation of relativistic cosmological models usually has
the energy momentum tensor of matter generated by a perfect fluid.
To consider more realistic models one must take into account the
viscosity mechanisms, which have already attracted the attention
of many researchers. Misner \cite{mis1,mis2} suggested that strong
dissipative due to the neutrino viscosity may considerably reduce
the anisotropy of the blackbody radiation. Viscosity mechanism in
cosmology can explain the anomalously high entropy per baryon in
the present universe \cite{wein,weinb}. Bulk viscosity associated
with the grand-unified-theory phase transition \cite{lang} may
lead to an inflationary scenario \cite{waga,pacher,guth}.

A uniform cosmological model filled with fluid which possesses pressure
and second (bulk) viscosity was developed by Murphy \cite{murphy}. The
solutions that he found exhibit an interesting feature that the big bang
type singularity appears in the infinite past. Exact solutions of the
isotropic homogeneous cosmology for open, closed and flat universe have
been found by Santos et al \cite{santos}, with the bulk viscosity being
a power function of energy density.

The nature of cosmological solutions for homogeneous Bianchi type
I (BI) model was investigated by Belinsky and Khalatnikov
\cite{belin} by taking into account dissipative process due to
viscosity. They showed that viscosity cannot remove the
cosmological singularity but results in a qualitatively new
behavior of the solutions near singularity. They found the
remarkable property that during the time of the \textit{big bang}
matter is created by the gravitational field. BI solutions in case
of stiff matter with a shear viscosity being the power function of
of energy density were obtained by Banerjee \cite{baner}, whereas
BI models with bulk viscosity ($\eta$) that is a power function of
energy density $\ve$ and when the universe is filled with stiff
matter were studied by Huang \cite{huang}. The effect of bulk
viscosity, with a time varying bulk viscous coefficient, on the
evolution of isotropic FRW models was investigated in the context
of open thermodynamics system was studied by Desikan
\cite{desikan}. This study was further developed by Krori and
Mukherjee \cite{krori} for anisotropic Bianchi models.
Cosmological solutions with nonlinear bulk viscosity were obtained
in \cite{chim}. Models with both shear and bulk viscosity were
investigated in \cite{elst,meln}.

Though Murphy \cite{murphy} claimed that the introduction of bulk
viscosity can avoid the initial singularity at finite past,
results obtained in \cite{barrow} show that, it is, in general,
not valid, since for some cases big bang singularity occurs in
finite past.

We studied a self-consistent system of the nonlinear spinor and/or
scalar fields in a BI spacetime  in presence of a perfect fluid
and a $\Lambda$ term \cite{sahaprd,bited} in order to clarify
whether the presence of a singular point is an inherent property
of the relativistic cosmological models or it is only a
consequence of specific simplifying assumptions underlying these
models. Recently we have considered a nonlinear spinor field in a
BI Universe filled with viscous fluid \cite{romrep}. Since the
viscous fluid itself presents a growing interest, we study the
influence of viscous fluid and $\Lambda$ term in the evolution of
the BI Universe in this report.

        \section{Derivation of Basic Equations}
Using the variational principle in this section we derive the
fundamental equations for the gravitational field  from the action
\eqref{action} :
\begin{equation}
{\cal S}(g; \ve) = \int\, \cL \sqrt{-g} d\Omega \label{action}
\end{equation}
with
\begin{equation}
\cL= \cL_{\rm grav.} + \cL_{\rm vf}. \label{lag}
\end{equation}

The gravitational part of the Lagrangian \eqref{lag} ${\cL}_{\rm
grav.}$ is given by a Bianchi type-I metric, whereas the term
${\cL}_{\rm vf}$ describes a viscous fluid.

We also write the expressions for the metric functions explicitly
in terms of the volume scale $\tau$ defined bellow \eqref{taudef}.
Defining Hubble constant \eqref{hubc} in analogy with a flat
Friedmann-Robertson-Walker (FRW) Universe, we also derive the
system of equations for $\tau$, $H$ and $\ve$, with $\ve$ being
the energy density of the viscous fluid, which plays the central
role here.

             \subsection{The gravitational field}
As a gravitational field we consider the Bianchi type I (BI) cosmological
model. It is the simplest model of anisotropic universe that describes
a homogeneous and spatially flat space-time and if filled with perfect
fluid with the equation of state $p = \zeta \ve, \quad \zeta < 1$, it
eventually evolves into a FRW universe \cite{jacobs}. The isotropy of
present-day universe makes BI model a prime candidate for studying the
possible effects of an anisotropy in the early universe on modern-day
data observations. In view of what has been mentioned above we choose
the gravitational part of the Lagrangian \eqref{lag} in the form
\begin{equation}
{\cal L}_{\rm g} = \frac{R}{2\kappa},
\label{lgrav}
\end{equation}
where $R$ is the scalar curvature, $\kappa = 8 \pi G$
being the Einstein's gravitational constant. The gravitational field in
our case is given by a Bianchi type I (BI) metric
\begin{equation}
ds^2 = dt^2 - a^2 dx^2 - b^2 dy^2 - c^2 dz^2,
\label{BI}
\end{equation}
with $a,\, b,\, c$ being the functions of time $t$ only. Here the speed of
light is taken to be unity.

The metric \eqref{BI} has the following non-trivial Christoffel symbols
\begin{eqnarray}
\G^{1}_{10} &=& \frac{\dot a}{a}, \quad  \G^{2}_{20} = \frac{\dot b}{b},
\quad \G^{3}_{30} = \frac{\dot c}{c} \nonumber \\
\label{chrsym}\\
\G^{0}_{11} &=& a {\dot a}, \quad
\G^{0}_{22} = b {\dot b}, \quad \G^{0}_{33} = c {\dot c}. \nonumber
\end{eqnarray}
The nontrivial components of the Ricci tensors are
\begin{subequations}
\label{ricten}
\begin{eqnarray}
R_{0}^{0} &=& - \Bigl(\frac{\ddot a}{a}+\frac{\ddot b}{b}+\frac{\ddot c}{c}
\Bigr), \label{rt00} \\
R_{1}^{1} &=& - \Bigl[\frac{\ddot a}{a} + \frac{\dot a}{a}\Bigl(
\frac{\dot b}{b} + \frac{\dot c}{c}\Bigr)\Bigr], \label{rt11} \\
R_{2}^{2} &=& - \Bigl[\frac{\ddot b}{b} + \frac{\dot b}{b}\Bigl(
\frac{\dot c}{c} + \frac{\dot a}{a}\Bigr)\Bigr], \label{rt22}\\
R_{3}^{3} &=& - \Bigl[\frac{\ddot c}{c} + \frac{\dot c}{c}\Bigl(
\frac{\dot a}{a} + \frac{\dot b}{b}\Bigr)\Bigr]. \label{rt33}
\end{eqnarray}
\end{subequations}
From \eqref{ricten} one finds the following Ricci scalar for the BI universe
\begin{equation}
R = - 2\Bigl(\frac{\ddot a}{a} + \frac{\ddot b}{b} + \frac{\ddot c}{c} +
\frac{\dot a}{a}\frac{\dot b}{b} + \frac{\dot b}{b}\frac{\dot c}{c}+
\frac{\dot c}{c}\frac{\dot a}{a}\Bigr).
\label{ricsc}
\end{equation}
The non-trivial components of Riemann tensors in this case read
\begin{eqnarray}
R^{01}_{\,\,\,\,01} &=& -\frac{\ddot a}{a}, \quad
R^{02}_{\,\,\,\,02} = -\frac{\ddot b}{b}, \quad
R^{03}_{\,\,\,\,03} = -\frac{\ddot c}{c}, \nonumber \\
\label{riemann} \\
R^{12}_{\,\,\,\,12} &=& -\frac{\dot a}{a}\frac{\dot b}{b}, \quad
R^{23}_{\,\,\,\,23} = -\frac{\dot b}{b}\frac{\dot c}{c}, \quad
R^{31}_{\,\,\,\,31} = -\frac{\dot c}{c}\frac{\dot a}{a}. \nonumber
\end{eqnarray}
Now having all the non-trivial components of Ricci and Riemann tensors,
one can easily write the invariants of gravitational field which we need
to study the space-time singularity. We return to this study at the end
of this section.

\subsection{Viscous fluid}

The influence of the viscous fluid in the evolution of the
Universe is performed by means of its energy momentum tensor,
which acts as the source of the corresponding gravitational field.
The reason for writing $\cL_{\rm vf}$ in \eqref{lag} is to
underline that we are dealing with a self-consistent system. The
energy momentum tensor of a viscous field has the form
\begin{equation}
T_{\mu\,{\rm (m)}}^{\nu} = (\ve + p^\prm) u_\mu u^\nu - p^{\prm}
\delta_\mu^\nu + \eta g^{\nu \beta} [u_{\mu;\beta}+u_{\beta:\mu}
-u_\mu u^\alpha u_{\beta;\alpha} - u_\beta u^\alpha
u_{\mu;\alpha}], \label{imper}
\end{equation}
where
\begin{equation}
p^{\prm} = p - (\xi - \frac{2}{3} \eta) u^\mu_{;\mu}. \label{ppr}
\end{equation}
Here $\ve$ is the energy density, $p$ - pressure, $\eta$ and $\xi$
are the coefficients of shear and bulk viscosity, respectively.
Note that the bulk and shear viscosities, $\eta$ and $\xi$, are
both positively definite, i.e.,
\begin{equation}
\eta > 0, \quad \xi > 0.
\end{equation}
They may be either constant or function of time or energy, such as
:
\begin{equation}
\eta = |A| \ve^{\alpha}, \quad \xi = |B| \ve^{\beta}.
\label{etaxi}
\end{equation}
The pressure $p$ is connected to the energy density by means of a
equation of state. In this report we consider the one describing a
perfect fluid :
\begin{equation}
p = \zeta \ve, \quad \zeta \in (0, 1]. \label{pzeta}
\end{equation}
Note that here $\zeta \ne 0$, since for dust pressure, hence
temperature is zero, that results in vanishing viscosity.

In a comoving system of reference such that $u^\mu =
(1,\,0,\,0,\,0)$ we have
\begin{subequations}
\label{total}
\begin{eqnarray}
T_{0\,{\rm (m)}}^{0} &=& \ve, \\
T_{1\,{\rm (m)}}^{1} &=& - p^{\prm} + 2 \eta \frac{\dot a}{a}, \\
T_{2\,{\rm (m)}}^{2} &=& - p^{\prm} + 2 \eta \frac{\dot b}{b}, \\
T_{3\,{\rm (m)}}^{3} &=& - p^{\prm} + 2 \eta \frac{\dot c}{c}.
\end{eqnarray}
\end{subequations}

Let us introduce the dynamical scalars such as the expansion and
the shear scalar as usual
\begin{equation}
\theta = u^\mu_{;\mu},\quad \sigma^2 = \frac{1}{2} \sigma_{\mu\nu}
\sigma^{\mu\nu}, \label{dynscal}
\end{equation}
where
\begin{equation}
\sigma_{\mu\nu} = \frac{1}{2} \Bigl(u_{\mu;\alpha}
P^{\alpha}_{\nu} + u_{\nu;\alpha} P^{\alpha}_{\mu}\Bigr) -
\frac{1}{3} \theta P_{\mu \nu}. \label{shearten}
\end{equation}
Here $P$ is the projection operator obeying
\begin{equation}
P^2 = P.
\end{equation}
For the space-time with signature $(+, \,-,\,-,\,-)$ it has the
form
\begin{equation}
P_{\mu\nu} = g_{\mu\nu} - u_\mu u_\nu, \quad P^\mu_\nu =
\delta^\mu_\nu - u^\mu u_\nu. \label{projec}
\end{equation}
For the BI metric the dynamical scalar has the form
\begin{equation}
\theta = \frac{\dot {a}}{a}+\frac{\dot {b}}{b} + \frac{\dot
{c}}{c} = \frac{\dot {\tau}}{\tau}, \label{expan}
\end{equation}
and
\begin{equation}
2 \sigma^2 = \frac{\dot {a}^2}{a^2}+\frac{\dot {b}^2}{b^2} +
\frac{\dot {c}^2}{c^2} - \frac{1}{3} \theta^2. \label{shesc}
\end{equation}

\subsection{Field equations and their solutions}

Variation of \eqref{action} with respect to metric tensor
$g_{\mu\nu}$ gives the Einstein's field equation. In account of
the $\Lambda$-term we then have
\begin{equation}
G_\mu^\nu = R_{\mu}^{\nu} - \frac{1}{2} \delta_{\mu}^{\nu} R =
\kappa T_{\mu}^{\nu} - \delta_{\mu}^{\nu} \Lambda.
\label{Ein}
\end{equation}
In view of \eqref{ricten} and \eqref{ricsc} for the BI space-time \eqref{BI}
we rewrite the Eq. \eqref{Ein} as
\begin{subequations}
\label{BID}
\begin{eqnarray}
\frac{\ddot b}{b} +\frac{\ddot c}{c} + \frac{\dot b}{b}\frac{\dot
c}{c}&=&  \kappa T_{1}^{1} -\Lambda,\label{11}\\
\frac{\ddot c}{c} +\frac{\ddot a}{a} + \frac{\dot c}{c}\frac{\dot
a}{a}&=&  \kappa T_{2}^{2} - \Lambda,\label{22}\\
\frac{\ddot a}{a} +\frac{\ddot b}{b} + \frac{\dot a}{a}\frac{\dot
b}{b}&=&  \kappa T_{3}^{3} - \Lambda,\label{33}\\
\frac{\dot a}{a}\frac{\dot b}{b} +\frac{\dot b}{b}\frac{\dot c}{c}
+\frac{\dot c}{c}\frac{\dot a}{a}&=&  \kappa T_{0}^{0} - \Lambda,
\label{00}
\end{eqnarray}
\end{subequations}
where over dot means differentiation with respect to $t$ and
$T_{\nu}^{\mu}$ is the energy-momentum tensor of a viscous fluid
given above \eqref{total}.

\subsubsection{Expressions for the metric functions }

To write the metric functions explicitly, we define a new time
dependent function $\tau (t)$
\begin{equation}
\tau = a b c = \sqrt{-g},
\label{taudef}
\end{equation}
which is indeed the volume scale of the BI space-time.

Let us now solve the Einstein equations.
In account of \eqref{total} subtracting \eqref{11} from \eqref{22},
one finds the following relation between $a$ and $b$
\begin{equation}
\frac{a}{b}= D_1 \exp \biggl(X_1 \int \frac{ e^{-2 \kappa \int
\eta dt}}{\tau}\,dt\biggr). \label{ab}
\end{equation}
Analogically, we find
\begin{eqnarray}
\frac{b}{c}= D_2 \exp \biggl(X_2 \int \frac{ e^{-2 \kappa \int
\eta dt}}{\tau}\,dt\biggr), \quad \frac{c}{a}= D_3 \exp \biggl(X_3
\int \frac{ e^{-2 \kappa \int \eta dt}}{\tau}\,dt\biggr).
\label{ac}
\end{eqnarray}
Here $D_1,\,D_2,\,D_3,\,X_1,\, X_2, X_3 $ are integration constants, obeying
\begin{eqnarray}
D_1 D_2 D_3 = 1, \quad X_1 + X_2 + X_3 = 0.
\label{intcon}
\end{eqnarray}

In view of \eqref{intcon} from \eqref{ab} and \eqref{ac}
we write the metric functions explicitly~\cite{sahaprd}
\begin{subequations}
\label{abc}
\begin{eqnarray}
a(t) &=& A_{1} \tau^{1/3}\exp\biggl[(B_1/3) \int\, \frac{ e^{-2
\kappa \int \eta dt}}{\tau}\,dt \biggr], \label{a} \\
b(t) &=& A_{2} \tau^{1/3}\exp\biggl[(B_2/3) \int\, \frac{ e^{-2
\kappa \int \eta dt}}{\tau}\,dt \biggr], \label{b}\\
c(t) &=& A_{3} \tau^{1/3}\exp\biggl[(B_3/3) \int\, \frac{ e^{-2
\kappa \int \eta dt}}{\tau}\,dt\biggr], \label{c}
\end{eqnarray}
\end{subequations}
where
\begin{eqnarray}
A_1 &=& \sqrt[3]{(D_1/D_3)},\quad A_2 = \sqrt[3]{1/(D_1^2
D_3)},\quad \sqrt[3]{(D_1 D_3^2)}, \nonumber\\
B_1 &=& X_1 - X_3, \quad B_2 = - (2 X_1 + X_3), \quad B_3 = X_1 +
2 X_3. \nonumber
\end{eqnarray}
Thus, the metric functions are found explicitly in terms of $\tau$
and viscosity.

As one sees from \eqref{a}, \eqref{b} and \eqref{c}, for $\tau = t^n$
with $n > 1$ the exponent tends to unity at large $t$, and the
anisotropic model becomes isotropic one.

\subsubsection{Singularity analysis}

Let us now investigate the existence of singularity (singular
point) of the gravitational case, which can be done by
investigating the invariant characteristics of the space-time. In
general relativity these invariants are composed from the
curvature tensor and the metric one. In a 4D Riemann space-time
there are 14 independent invariants. Instead of analyzing all 14
invariants, one can confine this study only in 3, namely the
scalar curvature $I_1 = R$, $I_2 = R_{\mu\nu}R^{\mu\nu}$, and the
Kretschmann scalar $I_3 =
R_{\alpha\beta\mu\nu}R^{\alpha\beta\mu\nu}$ \cite{bronshik,fay}.
At any regular space-time point, these three invariants
$I_1,\,I_2,\,I_3$ should be finite. Let us rewrite these
invariants in detail.

For the Bianchi I metric one finds the scalar curvature
\begin{eqnarray}
I_1 = R = - 2\Bigl(\frac{\ddot a}{a}+\frac{\ddot b}{b}+\frac{\ddot c}{c}+
\frac{\dot a}{a}\frac{\dot b}{b} + \frac{\dot b}{b}\frac{\dot c}{c}+
\frac{\dot c}{c}\frac{\dot a}{a}\Bigr).
\label{SC}
\end{eqnarray}
Since the Ricci tensor for the BI metric is diagonal, the
invariant $I_2 = R_{\mu\nu}R^{\mu\nu} \equiv R_{\mu}^{\nu} R_{\nu}^{\mu}$
is a sum of squares of diagonal components of Ricci tensor, i.e.,
\begin{equation}
I_2 = \Bigl[\bigl(R_{0}^{0}\bigr)^2  + \bigl(R_{1}^{1}\bigr)^2 +
\bigl(R_{2}^{2}\bigr)^2 + \bigl(R_{3}^{3}\bigr)^2 \Bigr],
\end{equation}
with the components of the Ricci tensor being given by \eqref{ricten}.

Analogically, for the Kretschmann scalar in this case we have
$I_3 = R_{\,\,\,\,\,\,\,\alpha\beta}^{\mu\nu}
R_{\,\,\,\,\,\,\,\mu\nu}^{\alpha\beta}$,
a sum of squared components of all nontrivial
$R_{\,\,\,\,\,\,\,\mu\nu}^{\mu\nu}$, which in view of \eqref{riemann}
can be written as
\begin{eqnarray}
I_3 &=& 4 \Biggl[
 \Bigl(R_{\,\,\,\,\,\,01}^{01}\Bigr)^2
+ \Bigl(R_{\,\,\,\,\,\,02}^{02}\Bigr)^2
+ \Bigl(R_{\,\,\,\,\,\,03}^{03}\Bigr)^2
+ \Bigl(R_{\,\,\,\,\,\,12}^{12}\Bigr)^2
+ \Bigl(R_{\,\,\,\,\,\,23}^{23}\Bigr)^2
+ \Bigl(R_{\,\,\,\,\,\,31}^{31}\Bigr)^2\Biggr] \nonumber\\
&=& 4\Bigl[\Bigl(\frac{\ddot a}{a}\Bigr)^2 +
\Bigl(\frac{\ddot b}{b}\Bigr)^2+\Bigl(\frac{\ddot c}{c}\Bigr)^2
+ \Bigl(\frac{\dot a}{a}\frac{\dot b}{b}\Bigr)^2 +
\Bigl(\frac{\dot b}{b}\frac{\dot c}{c}\Bigr)^2 +
\Bigl(\frac{\dot c}{c}\frac{\dot a}{a}\Bigr)^2\Bigr].
\label{Kretsch}
\end{eqnarray}
Let us now express the foregoing invariants in terms of $\tau$.
From Eqs. \eqref{abc} we have
\begin{subequations}
\label{sing}
\begin{eqnarray}
a_i &=& A_i \tau^{1/3} \exp \Biggl((B_i/3) \int\,\frac{e^{-2
\kappa \int \eta dt}}{\tau (t)} dt\Biggr),
\\
\frac{{\dot a}_i}{a_i} &=& \frac{\dot{\tau} + B_i e^{-2 \kappa
\int \eta dt}}{3 \tau}
\quad (i=1,2,3,),\\
\frac{{\ddot a}_i}{a_i} &=& \frac{3 \tau \ddot{\tau} - 2
\dot{\tau}^2 - \dot{\tau} B_i e^{-2 \kappa \int \eta dt} - 6
\kappa \eta \tau B_i e^{-2 \kappa \int \eta dt} + B_i^2 e^{-4
\kappa \int \eta dt}} {9 \tau^2},
\end{eqnarray}
\end{subequations}
i.e., the metric functions $a, b, c$ and their derivatives are in
functional dependence with $\tau$. From Eqs. \eqref{sing} one can easily
verify that
$$I_1 \propto \frac{1}{\tau^2},\quad
I_2 \propto \frac{1}{\tau^4},\quad I_3 \propto \frac{1}{\tau^4}.$$
Thus we see that at any space-time point, where $\tau = 0$ the
invariants $I_1,\,I_2,$ and $I_3$ become infinity, hence the
space-time becomes singular at this point.

\subsection{Equations for determining $\tau$}

In the foregoing subsection we wrote the corresponding metric
functions in terms of volume scale $\tau$. In what follows, we
write the equation for $\tau$ and study it in details.

Summation of Einstein equations \eqref{11}, \eqref{22}, \eqref{33}
and 3 times \eqref{00} gives
\begin{equation}
{\ddot \tau} - \frac{3}{2} \kappa \xi {\dot \tau}
= \frac{3}{2}\kappa \bigl(\ve - p \bigr) \tau - 3 \Lambda \tau.
\label{dtau1}
\end{equation}
For the right-hand-side of \eqref{dtau1} to be a function
of $\tau$ only, the solution to this equation is well-known~\cite{kamke}.

Let us demand the energy-momentum to be conserved, i.e.,
\begin{equation}
T_{\mu;\nu}^{\nu} = T_{\mu,\nu}^{\nu} + \G_{\rho\nu}^{\nu} T_{\mu}^{\rho}
- \G_{\mu\nu}^{\rho} T_{\rho}^{\nu} = 0,
\end{equation}
which in our case has the form
\begin{equation}
\frac{1}{\tau}\bigl(\tau T_0^0\bigr)^{\cdot} - \frac{\dot a}{a} T_1^1
-\frac{\dot b}{b} T_2^2  - \frac{\dot c}{c} T_3^3 = 0.
\label{emcon}
\end{equation}
After a little manipulation from
\eqref{emcon} we obtain
\begin{equation}
{\dot \ve} + \frac{\dot \tau}{\tau} \omega -
(\xi + \frac{4}{3} \eta) \frac{{\dot \tau}^2}{\tau^2} +
4 \eta (\kappa T_0^0 - \Lambda) = 0,
\label{vep}
\end{equation}
where
\begin{equation}
\omega = \ve + p,
\label{thermal}
\end{equation}
is the thermal function.

Let us now, in analogy with Hubble constant in a FRW Universe,
introduce a generalized Hubble constant $H$ :
\begin{equation}
\frac{\dot {\tau}}{\tau} = \frac{\dot {a}}{a}+\frac{\dot {b}}{b} +
\frac{\dot {c}}{c} = 3 H.
\label{hubc}
\end{equation}
Then \eqref{dtau1} and \eqref{vep} in account of \eqref{total} can be
rewritten as
\begin{subequations}
\label{HVe}
\begin{eqnarray}
\dot {H} &=& \frac{\kappa}{2}\bigl(3 \xi H - \omega\bigr) -
\bigl(3 H^2 - \kappa \ve + \Lambda \bigr) ,  \label{H}\\
\dot {\ve} &=& 3 H\bigl(3 \xi H - \omega\bigr) + 4 \eta \bigl(3
H^2 - \kappa \ve + \Lambda \bigr) . \label{Ve}
\end{eqnarray}
\end{subequations}

The Eqs. \eqref{HVe} can be written in terms of dynamical scalar
as well.

In account of \eqref{abc} one can also rewrite share scalar
\eqref{shesc} as
\begin{equation}
2 \sigma^2 = \frac{6 (X_1^2 + X_1 X_3 + X_3^2)}{9 \tau^2} e^{-4
\kappa \int \eta dt}.
\end{equation}
From \eqref{00} now yields
\begin{equation}
\frac{1}{3} \theta^2 - \sigma^2 = \kappa  \ve - \Lambda
\label{sh00}
\end{equation}
The Eqs. \eqref{HVe} now can be written in terms of $\theta$ and
$\sigma$ as follows
\begin{subequations}
\label{TS}
\begin{eqnarray}
\dot {\theta} &=& \frac{3\kappa}{2}\bigl(\xi \theta - \omega\bigr) -
3 \sigma^2,  \label{theta}\\
\dot {\ve} &=& \theta \bigl(\xi \theta - \omega\bigr) + 4 \eta \sigma^2.
\label{Vsig}
\end{eqnarray}
\end{subequations}
Note that the Eqs. \eqref{TS} coincide with the ones given in
\cite{baner}.

 \subsection{Some special solutions}

In this subsection we simultaneously solve the system of equations
for $\tau$, $H$, and $\ve$. It is convenient to rewrite the Eqs.
\eqref{hubc} and \eqref{HVe} as a single system :
\begin{subequations}
\label{HVen}
\begin{eqnarray}
\dot \tau &=& 3 H \tau, \label{tauH}\\
\dot {H} &=& \frac{\kappa}{2}\bigl(3 \xi H - \omega\bigr) -
\bigl(3 H^2 - \kappa \ve + \Lambda \bigr) ,  \label{Hn}\\
\dot {\ve} &=& 3 H\bigl(3 \xi H - \omega\bigr) + 4 \eta \bigl(3
H^2 - \kappa \ve + \Lambda \bigr) . \label{Ven}
\end{eqnarray}
\end{subequations}

In account of \eqref{thermal},\eqref{etaxi} and \eqref{pzeta} the
Eqs. \eqref{HVen} now can be rewritten as
\begin{subequations}
\label{HVen1}
\begin{eqnarray}
\dot \tau &=& 3 H \tau, \label{tauH1}\\
\dot {H} &=& \frac{\kappa}{2}\bigl(3 B \ve^\beta H - (1+\zeta)\ve
\bigr) - \bigl(3 H^2 - \kappa\ve + \Lambda \bigr) ,
\label{Hn1}\\
\dot {\ve} &=& 3 H\bigl(3 B \ve^\beta H - (1+\zeta)\ve \bigr) + 4
A \ve^\alpha \bigl(3 H^2 -  \kappa \ve + \Lambda \bigr) .
\label{Ven1}
\end{eqnarray}
\end{subequations}

The system \eqref{HVen} have been extensively studied in
literature either partially \cite{murphy,huang,baner} or in
general \cite{belin}. In what follows, we consider the system
\eqref{HVen} for some special choices of the parameters.

        \subsubsection{Case with bulk viscosity}
Let us first consider the case when the real fluid possesses the
bulk viscosity only. The corresponding system of Eqs. can then be
obtained by setting $\eta = 0$ in \eqref{HVen} or $A = 0$ in
\eqref{HVen1}. In this case the Eqs. \eqref{tauH} and \eqref{Hn}
remain unaltered, while \eqref{Ven} takes the form
\begin{equation}
\dot {\ve} = 3 H\bigl(3 \xi H - \omega\bigr). \label{vetau}
\end{equation}
In view of \eqref{vetau} the system \eqref{HVen} admits the
following first integral
\begin{equation}
\tau^2 \bigl(\kappa \ve - 3 H^2 - \Lambda \bigr) = C_1, \quad C_1
= {\rm const.} \label{int1}
\end{equation}
The relation \eqref{int1} can be interpreted as follows. At the
initial stage of evolution the volume scale $\tau$ tends to zero,
while, the energy density $\ve$ tends to infinity. Since the
Hubble constant and the $\Lambda$ term are finite, the relation
\eqref{int1} is in correspondence with the current line of
thinking. Let us see what happens as the Universe expands. It is
well known that with the expansion of the Universe, i.e., with the
increase of $\tau$, the energy density $\ve$ decreases. Suppose at
some stage of expansion $\tau \to \infty$ and $\ve \to 0$. Then
from \eqref{int1} follows that at the stage in question
\begin{equation}
3 H^2 + \Lambda \to 0. \label{asympan}
\end{equation}
In case of $\Lambda = 0$, we find $H = 0$, i.e., in absence of a
$\Lambda$ term, once $\tau \to \infty$, the process of evolution
is terminated. As one sees from \eqref{asympan}, for the $H$ to
make any sense, the $\Lambda$ term should be negative. In presence
of a negative $\lambda$ term the evolution process of the Universe
never comes to a halt, it either expands further or begin to
contract depending on the sign of $H = \pm \sqrt{- \Lambda/3}$,
\quad $\Lambda < 0$. Thus we see that the Universe may be
infinitely large only if $\Lambda \le 0$.

Let us now consider the case when the bulk viscosity is inverse
proportional to expansion, i.e.,
\begin{equation}
\xi \theta = C_2, \quad C_2 = {\rm const.} \label{xthinv}
\end{equation}
Now keeping into mind that $\theta = {\dot \tau}/\tau = 3 H$, also
the relations \eqref{tauH}, \eqref{thermal} and \eqref{pzeta} the
Eq. \eqref{vetau} can be written as
\begin{equation}
\frac{\dot \ve}{C_2 - (1 + \zeta) \ve} = \frac{\dot \tau}{\tau}.
\label{vetau1}
\end{equation}
From the Eq. \eqref{vetau1} one finds
\begin{equation}
\ve = \frac{1}{1 + \zeta} \bigl[C_2 + C_3 \tau^{-(1 +
\zeta)}\bigr], \label{vetauex}
\end{equation}
with $C_3$ being some arbitrary constant. Further, inserting $\ve$
from \eqref{vetauex} into \eqref{dtau1} one finds the expression
for $\tau$ explicitly.

Taking into account the equation of state \eqref{pzeta} in view of
\eqref{xthinv} and \eqref{vetauex}, the Eq. \eqref{dtau1} admits
the following solution in quadrature :
\begin{equation}
\int \frac{d \tau}{\sqrt{C_2^2 + C_0^0 \tau^2 + C_1^1 \tau^{1 -
\zeta}}} = t + t_0, \label{quadr}
\end{equation}
where $C_2^2$ and $t_0$ are some constants. Further we set $t_0 =
0$. Here, $C_0^0 = 3 \kappa C_2/(1 + \zeta) - 3 \Lambda$ and
$C_1^1 = 3 \kappa C_3 /(1 + \zeta)$. As one sees, $C_0^0$ is
negative for
\begin{equation}
\Lambda > \kappa C_2/(1 + \zeta). \label{Lamp}
\end{equation}
It means that for a positive $\Lambda$ obeying \eqref{Lamp} (we
assume that the constant $C_2$ is a positive quantity) $\tau$
should be bound from above as well. Let us now rewrite the Eq.
\eqref{quadr} in the form
\begin{equation}
\dot{\tau} = \sqrt{2[E - {\cal U}(\tau)]}, \label{1stint}
\end{equation}
where $E = C_2^2/2$ can be viewed as energy and ${\cal U}(\tau) =
-0.5(C_0^0 \tau^2 + C_1^1 \tau^{1- \zeta})$ can be seen as the
potential [cf. Fig. \ref{poten}] corresponding to the Eqn.
\eqref{dtau1}. Depending on the value of $E$ there exists two
types of solutions: for $E > 0$ we have non-periodic solutions,
i.e., after reaching some maximum value (say $\tau_{\rm max}$) the
BI Universe begin to contract and finally collops into a point,
thus giving rise to a space-time singularity; for $E < 0$ BI
space-time admits a singularity-free oscillatory mode of expansion
[cf. Fig. \ref{oscl}]. A comprehensive description concerning
potential can be found in \cite{bited}.

\myfigures{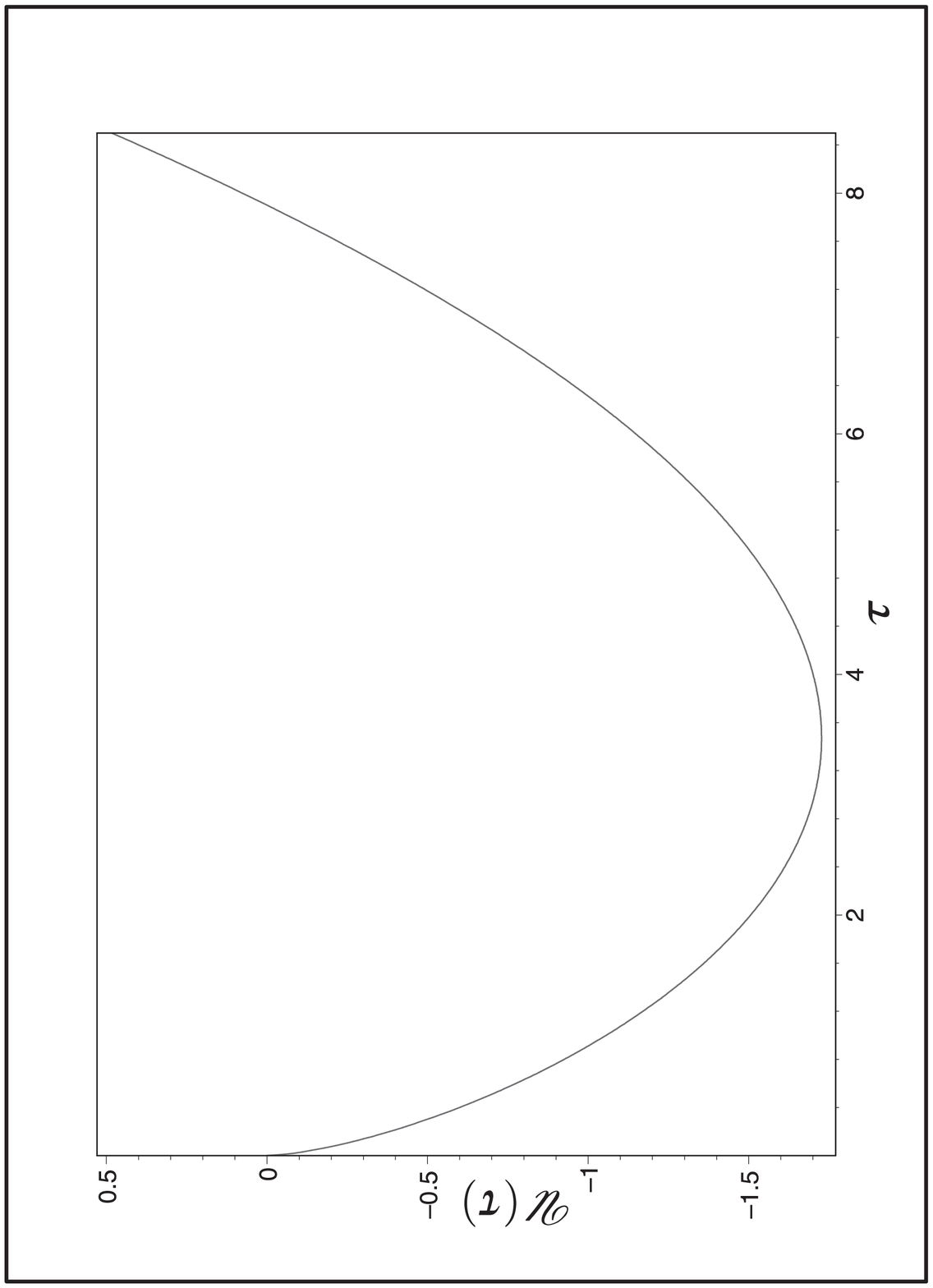}{0.35}{View of the potential ${\cal U}(\tau)$.
Here we set $\kappa = 1$, $C_2 = 1$, and $C_3 = 1$. Perfect fluid
corresponds to a radiation, i.e., $\zeta = 0.33$, and the
Cosmological constant is taken to be $\Lambda = 0.8$
}{0.45}{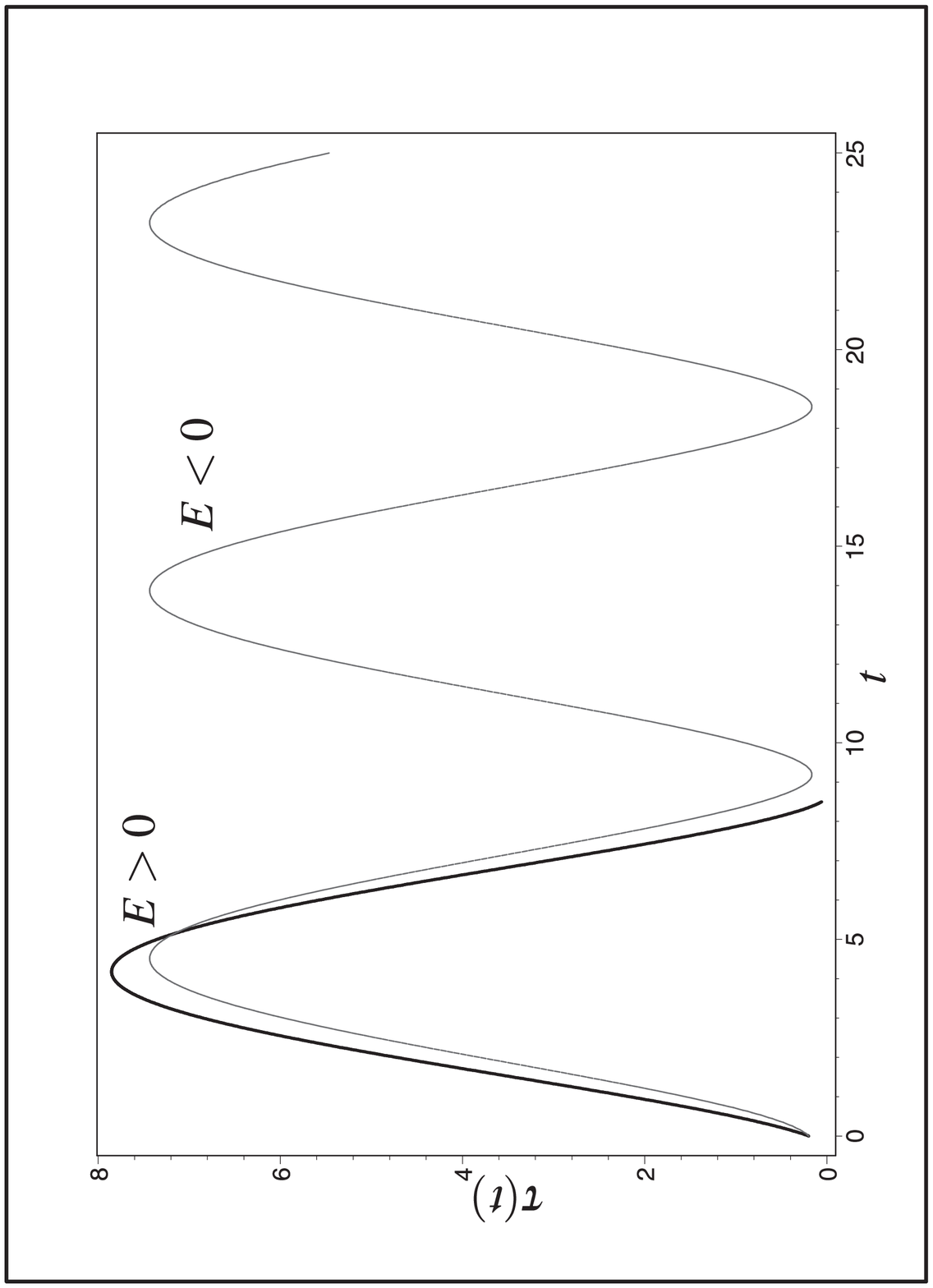}{0.35}{Evolution of the BI space-time corresponding
to the potential given in Fig. \ref{poten}. The initial value of
the volume scale in this case is taken to be $\tau_0 =
0.2$.}{0.45}

As a second example we consider the case, when $\zeta = 1$. From
\eqref{quadr} one then finds
\begin{subequations}
\label{exposc}
\begin{eqnarray}
\tau (t) &=& \bigl(\exp(\sqrt{C_0^0}\,\,t) - C_2^2
\exp(\sqrt{C_0^0}\,\,t)\bigr)/(2 \sqrt{C_0^0}), \quad C_0^0 > 0, \label{exp} \\
\tau (t) &=& (C_2^2 / \sqrt{|C_0^0|}) \sin\,(\sqrt{|C_0^0|}\,\,
t). \quad C_0^0 < 0 \label{osc}.
\end{eqnarray}
\end{subequations}
Taking into account that $C_0^0 > 0$ for any non-positive
$\Lambda$, from \eqref{exp} one sees that, in case of $\Lambda \le
0$ the Universe may be infinitely large (there is no upper bound),
which is in line with the conclusion made above. On the other
hand, $C_0^0$ may be negative only for some positive value of
$\Lambda$. It was shown in \cite{sahaprd,bited} that in case of a
perfect fluid a positive $\Lambda$ always invokes oscillations in
the model, whereas, in the present model with viscous fluid, it is
the case only when $\Lambda$ obeys \eqref{Lamp}. Unlike the case
with radiation where BI admits two types of solutions, the case
with stiff matter allows only one type of solutions, namely the
non-periodic one that corresponds to $E > 0$ in the previous case,
since now potential ${\cal U}(\tau) =  -0.5 C_0^0 \tau^2$ has its
minimum at $\tau = 0$.

\subsubsection{Case with shear and bulk viscosity}

Let us now consider the general case with the shear viscosity
$\eta$ being proportional to the expansion, i.e.,
\begin{equation}
\eta \propto \theta = 3 H. \label{etath}
\end{equation}
We will consider the case when
\begin{equation}
\eta = -\frac{3}{2 \kappa} H. \label{etath1}
\end{equation}
In this case from \eqref{Hn} and \eqref{Ven} one easily find
\begin{equation}
3 H^2 = \kappa \ve + C_4, \quad C_4 = {\rm const.} \label{shv}
\end{equation}
From \eqref{shv} it follows that at the initial state of
expansion, when $\ve$ is large, the Hubble constant is also large
and with the expansion of the Universe $H$ decreases as does
$\ve$. Inserting the relation \eqref{shv} into the Eqs. \eqref{Hn}
one finds
\begin{equation}
\int \frac{d H}{\sqrt{A H^2 + B H + C}} = t, \label{quad1}
\end{equation}
where, $A = - 1.5 (1 + \zeta)$,\, $B = 1.5 \kappa \xi$, and $C =
0.5 C_4 (\zeta - 1) - \Lambda$. If the bulk viscosity is taken to
be a constant one, i.e., $\xi = {\rm const.}$, then in view of the
fact that $A < 0$, the Hubble constant $H$ admits sinusoidal form,
namely, \cite{prud}
\begin{equation}
H = - \frac{\sqrt{B^2 - 4 AC}\, \sin\,(\sqrt{-A}\,\,t) + B}{2 A},
\label{sinHub}
\end{equation}
if and only if
\begin{equation}
\sqrt{B^2 - 4 AC} = (9/4) \kappa^2 \xi^2 - 3 (1 - \zeta^2) C_4 - 6
(1 + \zeta) \Lambda > 0. \label{adcon}
\end{equation}
As one sees the inequality \eqref{adcon} can be attained with a
negative $\Lambda$ as well. Further, from \eqref{tauH} one finds
the expression for $\tau$:
\begin{equation}
\tau(t) = C_0 \exp\Bigl[-\frac{3}{2}\Bigl\{ \frac{-\sqrt{B^2-4 A
C}\,\cos(\sqrt{-a}\,t) + B \sqrt{-A}\,t}{A
\sqrt{-A}}\Bigr\}\Bigr]. \label{taubs}
\end{equation}
A graphical view of the evolution of $\tau$ is given in Fig.
\ref{osclbs}.

\myfigure{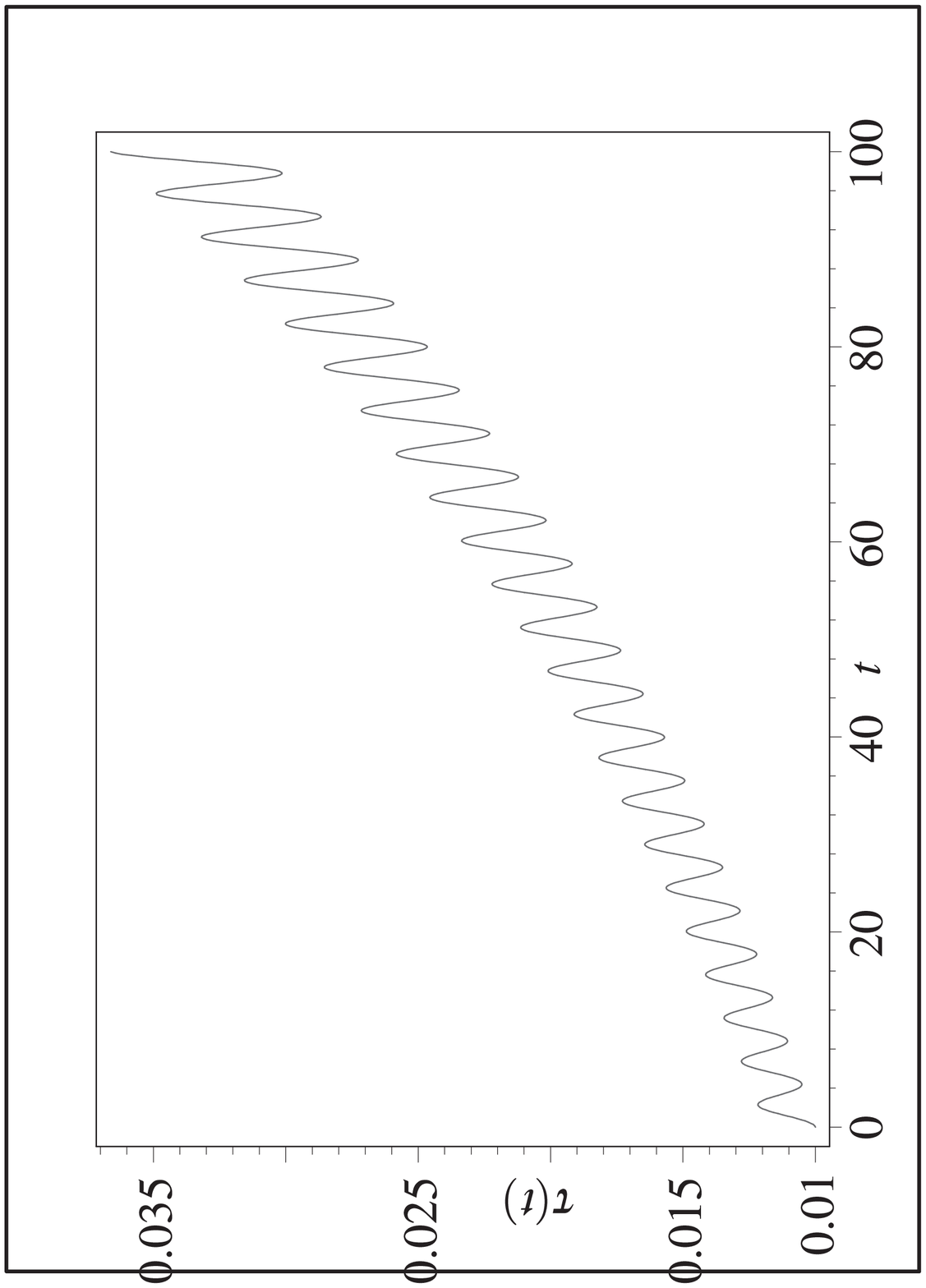}{0.30}{Evolution of the BI Universe with a bulk
viscosity and a share viscosity. The volume scale $\tau$ in this
case expands exponentially with a small oscillation.}{0.70}

As one sees from the Fig. \ref{osclbs} the exponential mode of
evolution of the BI Universe is accompanied by an oscillation. The
space-time singularity in this case arises at $t \to -\infty$.
Note that the negative $\Lambda$ gives rise to the exponential
growth while the oscillation in the model occurs due to the
viscosity.


               \section{Conclusion}
The role of viscous fluid and $\Lambda$ term in the evolution of a
homogeneous, anisotropic Universe given by a Bianchi type-I
space-time is studied. It is shown that the $\Lambda$ term plays
very important role in BI cosmology. In particular, in case of a
bulk viscosity, it provides an everlasting process of evolution
with $\Lambda$ being negative, whereas, for some positive values
of $\Lambda$ the BI Universe admits an oscillatory mode of
expansion. It is also shown that a oscillatory mode of expansion
of the BI space-time can take place with a negative $\Lambda$ as
well, though it is accompanied by an exponential mode. Oscillation
in this case arises due to viscosity. In this report we only
consider some special cases those provides exact solutions. For a
better knowledge about the evolution, it is important to perform
some qualitative analysis of the system \eqref{HVen}. A detailed
analysis of the system in question plus some numerical solutions
will be presented soon elsewhere.


\newcommand{\hnl}{\htmladdnormallink}


\begin{thebibliography}{99}


\bibitem{mis1} W. Misner, Nature {\bf 214}, 40 (1967).

\bibitem{mis2} W. Misner, Astrophys. J. {\bf 151}, 431 (1968).

\bibitem{wein} S. Weinberg, Astrophys. J. {\bf 168}, 175 (1972).

\bibitem{weinb} S. Weinberg, {\it Gravitation and Cosmology}
    (New York, Wiley, 1972).

\bibitem{lang} P. Langacker, Phys. rep. {\bf 72}, 185 (1981).

\bibitem{waga} L. Waga, R.C. Falcan, and R. Chanda, Phys. Rev. D
    {\bf 33}, 1839 (1986).

\bibitem{pacher} T. Pacher, J.A. Stein-Schabas, and M.S. Turner,
    Phys. Rev. D {\bf 36}, 1603 (1987).

\bibitem{guth} Alan Guth, Phys. Rev. D {\bf 23}, 347 (1981).

\bibitem{murphy} G.L. Murphy, Phys. Rev. D {\bf 8}, 4231 (1973).

\bibitem{santos} N.O. Santos, R.S. Dias, and A. Banerjee, J.
Math. Phys. {\bf 26}, 876 (1985).

\bibitem{belin} V.A. Belinski and I.M. Khalatnikov, Soviet Journal JETP
{\bf 69}, 401, (1975).

\bibitem{baner} A. Banerjee, S.B. Duttachoudhury, and A.K. Sanyal, J.
Math. Phys. {\bf 26}, 3010 (1985).

\bibitem{huang} W. Huang, J. Math. Phys. {\bf 31},
    1456 (1990).

\bibitem{desikan} K. Desikan, Gen. Relativ. Gravit. {\bf 29},
    435 (1997).

\bibitem{krori} K.D. Krori and A. Mukherjee, Gen. Relativ.
    Gravit. {\bf 32}, 1429 (2000).

\bibitem{chim} L.P. Chimento, A.S. Jacubi, V. M$\grave e$ndez, and R.
    Maartens, Class. Quantum Grav. {\bf 14}, 3363 (1997).

\bibitem{elst} H. van Elst, P. Dunsby, and R. Tavakol, Gen. Relativ.
     Gravit. {\bf 27}, 171 (1995).

\bibitem{meln} V.R. Gavrilov, V.N. Melnikov, and R. Triay,
    Class. Quantum Grav. {\bf 14}, 2203 (1997).

\bibitem{barrow} J.D. Barrow, Nuc. Phys. B {\bf 310}, 743 (1988).

\bibitem{sahaprd} Bijan Saha,
Phys. Rev. D {\bf 64},
\hnl{123501}{http://www.jinr.ru/~bijan/my_papers/PRD23501.pdf}
(2001).

\bibitem{bited} Bijan Saha,
Phys. Rev. D {\bf 69},
\hnl{124010}{http://www.jinr.ru/~bijan/my_papers/PRD24010.pdf}
(2004).

\bibitem{romrep} Bijan Saha,
"Nonlinear spinor field in Bianchi type-I Universe filled with
viscous fluid: some special
\hnl{solutions"}{http://www.jinr.ru/~bijan/my_pdf/visnls.pdf} (to
appear in Romanian Report of Physics in 2005).

\bibitem{jacobs} K.C. Jacobs, Astrophys. J. {\bf 153}, 661 (1968).

\bibitem{bronshik} K.A. Bronnikov and G.N. Shikin,
Gravitation Cosmol. {\bf 7}, 231 (2001).

\bibitem{fay} S. Fay, Class. Quantum Grav. {\bf 17}, 2663 (2000).

\bibitem{kamke} E. Kamke, {\it Differentialgleichungen
losungsmethoden und losungen} (Akademische Verlagsgesellschaft,
Leipzig, 1957).

\bibitem{prud} A.P. Prudnikov, Yu.A. Brychkov, and O.I. Marichev,
{\it Integrals and series. v.1: Elementary functions} (New York,
Gordon and Breach, 1986).
\end{thebibliography}
\end{document}